\documentclass[preprint,12pt]{elsarticle}
\usepackage[utf8]{inputenc}
\usepackage{lineno}
\usepackage{amssymb}
\usepackage{amsmath}
\usepackage{float} 
\journal{Nature Cities}
\begin{document}

\begin{flushleft}
{\Large
\textbf\newline{Emergent regularity of walking patterns: evidence from Swiss cities} 
}
\newline

Guo-Shiuan Lin\textsuperscript{1},
Martin Hendrick\textsuperscript{1},
Maider Llaguno-Munitxa\textsuperscript{2},
Gabriele Manoli\textsuperscript{1,*}
\\
\bigskip
\textbf{1} Laboratory of Urban and Environmental Systems, School of Architecture, Civil and Environmental Engineering, École Polytechnique Fédérale de Lausanne, Switzerland \\
\textbf{2} Louvain Research Institute of Landscape, Architecture, Built Environment, Université Catholique de Louvain (UCLouvain), Ottignies-Louvain-la-Neuve 1348, Belgium \\
* gabriele.manoli@epfl.ch

\end{flushleft}

\section*{Abstract}
Human travel behaviour is characterised by high temporal and spatial regularity, as demonstrated by the Exploration and Preferential Return (EPR) mechanism, whereby individuals alternate between exploring new locations and preferentially returning to previously visited places. Yet, existing evidence is based on aggregated mobility data that cannot distinguish short trips or separate active mobility (i.e., walking and biking) from other mobility modes. This limits our ability to assess the benefits of walking activities and proximity-based policies such as the 15-minute city. Here, we analyse the spatial distribution and duration of walking habits of 1,318 participants, tracked during a 4-week time use survey in major Swiss cities - which are very close to the 15-minute city model. We find that walking activity is highly concentrated around individuals' home and work locations. These two attractors create walking patterns that are fundamentally different from other transport modes, which are instead used for longer trips and to connect home-workplace. We then introduce a dual-center EPR model, with empirically derived parameters, to simulate the decaying probability of walking away from the centers. Combining empirical data and the EPR framework, we find that groups with higher walking times show a stronger tendency to take long-distance walks and explore more destinations. In general, walking behavior reflects both individual circumstances and residential context: walking durations are shorter among people without a public transport pass, car owners, and residents of low-density rural areas. These findings provide a basis for more targeted assessments of walkability and active-mobility-oriented urban design.

\section{Introduction}
Walking, the most basic form of human mobility and physical activity, provides numerous health benefits, such as lower cardiovascular risk, prevention of cognitive decline, reduced obesity rates, and lower mortality \cite{paluch_daily_2022, weuve_physical_2004, bassett_walking_2008, hamer_walking_2008}. However, one third of the global adult population does not walk or engage in physical activity for more than 150 minutes weekly \cite{strain_national_2024, guthold_worldwide_2018} - the threshold recommended by the World Health Organization \cite{noauthor_who_2020}. The built environment density, diversity, designs, distance to transit, and destination accessibility considered as the five main drivers, so called 5Ds \cite{saelens_built_2008, mccormack_search_2011, frank_linking_2005, cervero_travel_1997, ewing_travel_2010} largely affects our walking behavior with Therefore, walkability has been considered an important urban design criteria for healthy and sustainable cities \cite{lowe_city_2022, cerin_determining_2022}. This shift toward active transport from car-oriented urban planning is further promoted by the 15-minute city concept—a proximity-based strategy where essential services are accessible within a 15-minute walk or cycle from home \cite{bruno_universal_2024, moreno_introducing_2021}. Yet, promoting active mobility in diverse and heterogeneous urban environments requires a deeper understanding of human walking behaviour, particularly its spatial patterns and variability across populations.

Previous research found that the probability distribution of walking trips follow an exponential decay with walking distance \cite{yang_walking_2012}. In the US, most walks are below 600 meters with few above 1200 m \cite{yang_walking_2012, millward_active-transport_2013, iacono_measuring_2010}. As such, 400 meters has been generally regarded as the 'acceptable walking distance' in the US urban design guidelines - although there is a slight variation depending on trip purposes and destinations \cite{yang_walking_2012, millward_active-transport_2013}. Studies have found that the most common walking origins and destinations are homes (25 \%) and workplaces (19.3 \%) with the other major walking destinations being grocery stores, transit stops, and restaurants \cite{millward_active-transport_2013}. Considering the typical upper walking distance and locations, the 1000m walkable buffer around home is commonly used to study the effects of neighborhood environment on walking \cite{cerin_determining_2022, frank_linking_2005}. Although the above statistics are often used in urban planning policy, walking behavior can differ among population subgroups with different household sizes, incomes, education levels, and racial or ethnic backgrounds \cite{yang_walking_2012, agrawal_extent_2007, christie_cross-sectional_2022}. Ownership of cars and dogs also affects the walking time \cite{brown_relationships_2006,sehatzadeh_walking_2011}. People who have easy access to public transport or own public transport passes are more likely to walk for transit, which contributes largely towards total walking time \cite{besser_walking_2005, freeland_walking_2013, cerin_determining_2022, lachapelle_transit_2009}. In summary, walking behavior is shaped by a variety  of individual and environmental features \cite{adkins_contextualizing_2017, van_dyck_relationships_2011,lee_correlates_2006}. 

In recent years, the increasing availability of mobility data have provided new opportunities to elucidate hidden patterns of population movement. Commonly used data sources include Call Detail Records (CDR) \cite[e.g.][]{gonzalez_understanding_2008, song_modelling_2010}, geotags on social media \cite[e.g.][]{jurdak_understanding_2015,osorio-arjona_social_2019}, bike sharing system \cite[e.g.]{kon_abstracting_2022}, public transit cards \cite[e.g.][]{cats_identifying_2024, zhong_measuring_2015}, traffic counting sensors \cite[e.g.][]{carter_enhancing_2020}, mobile Apps \cite{althoff_countrywide_2025, althoff_large-scale_2017} and GPS tracking \cite{lin_mining_2014, pappalardo_returners_2015, rhee_levy-walk_2011}. Studies from diverse data sources have found that human mobility possesses high regularity and predictability  \cite{schlapfer_universal_2021, gonzalez_understanding_2008, song_limits_2010, jurdak_understanding_2015}. Spatially, individuals' trajectories follow an universal probability distribution \cite{gonzalez_understanding_2008}, which center around their two most-frequented locations: home and workplace \cite{wang_extended_2019}. Temporally, return periods to the same locations follow diurnal cycles \cite{gonzalez_understanding_2008, jurdak_understanding_2015, song_limits_2010} and waiting times (time spent staying at one location between trips) follow a exponential or lognormal distribution   \cite{wang_extended_2019, jin_urban-epr_2025}. 

Based on these underlying mechanisms, various models have been proposed to reproduce human mobility patterns. At population level, models such as the gravitational model \cite{zipf_p1_1946} and radiation model \cite{simini_universal_2012} simulate aggregated travel flows between origins and destinations. At individual level, models have evolved from random walks, such as Levy flights and Brownian motion, to incorporate more realistic human mobility behavior \cite{barbosa_human_2018}. One such model is the Exploration and Preferential Return (EPR) which considers the tendency to return to visited locations \cite{song_modelling_2010}. In the EPR model, individuals are broadly classified into two groups: 'explorers' who constantly visit new places and 'returners' whose mobility is dominated by a few anchors \cite{pappalardo_returners_2015}. The EPR model has the advantage of having few parameters and offering mechanistic simplicity. Since its introduction, temporal structures have been integrated into the EPR framework to produce even more realistic mobility behaviors: the TimeGeo modelling framework combines circadian rhythm with rank-based EPR \cite{jiang_timegeo_2016}; the Ditras (DIary-based TRAjectory Simulator) uses mobility diary generator \cite{pappalardo_data-driven_2018}; the STS-EPR further considers social influence mechanisms between individuals \cite{cornacchia_sts-epr_2021}. These models capture well the overall mobility patterns (i.e., not exclusive to a certain transport mode) and have been extensively validated with empirical data. 

Despite these advances in data availability and model development, there are still several limitations that hinder our understanding of mobility, especially for walking. The first limitation is related to transport modes as many datasets are mode-specific and do not account for walking activities (e.g., data from public transport check-ins, bike sharing systems, or ride-hailing Apps). The datasets that cover large spatial domains, such as CDR and social media, are also not able to accurately distinguish walking from other transport modes. Also, they tend to miss information on short trips (e.g., trips within the reception area of the same cell tower), which is the case for most walking trips \cite{zhao_understanding_2016, chen_enriching_2018,rhee_levy-walk_2011, lin_mining_2014}. The dataset that can separate walking from other modes, such as  automatic pedestrian counting stations, encounter the second challenge: spatial coverage. The counting stations usually cover only the main road sections in central urban areas (e.g., in Zurich \cite{stadt_zurich_open_2026}, Melbourne \cite{carter_enhancing_2020}, and Dublin \cite{dublin_city_council_pedestrian_2026}). The third constrain is on the limited information about individual's characteristics. Data from sources like counting stations provide aggregated values and lack information on the population compositions and personal characteristics. The same is true for mobile step tracking Apps which generally count steps but do not record detailed trajectories. These limitations have prevented previous studies from fully characterizing the spatial distribution and within-population variation of walking activities - thus hindering also the development of accurate active mobility models. 

To bridge this gap, this study analyzes mobility behavior across major Swiss cities using a 4-week time use survey \cite{winkler_timeuse_2024} with comprehensive information on participants and their daily trips, including detailed attributes on locations, distance, duration, modes of transport, purposes, etc. After isolating walking activities, we show that most walks are short trips around home and workplace, contrasting to driving and cycling which are used for connecting these two places. Based on this finding, we propose a dual-center Exploration and Preference Return (EPR) model to explain the emergence of the bimodal walking behaviour. Finally, we show that walking durations depend on both personal and environmental features which should be considered for planning inclusive 15-minute cities. By comparing empirical mobility observations with model simulations, we shed a new light on walking behavior and provide guidelines for a transition to healthier urban environments.

\section{Methods}
\subsection{Mobility and activity data}
We access the mobility and activity data from the TimeUse+ survey \cite{winkler_timeuse_2024}, which employed the TimeUse+ mobile phone App (incorporating MotionTag’s Software Development Kit) to track the movements and activities of 1,318 voluntary participants in Switzerland for four consecutive weeks (between July 2022 and February 2023). The App uses GPS to passively track participants and automatically detects transport modes which are validated by the participants. For more details on the tracking methods and the survey, please refer to Winkler et al. (2024) \cite{winkler_timeuse_2024}. Along with the GPS tracks, the dataset also records the start, mid, and end location of each trip at 100m resolution. We define the euclidean distance between walking end or mid point to home as $D_{\rm home, end}$ and $D_{\rm home, mid}$ respectively. For the working population, we define the euclidean distance between workplace and home as $D_{\rm home, work}$, which we use to normalize $D_{\rm home, end}$ and $D_{\rm home, mid}$ across the population. The normalized distance between walking end points and home is defined as $\tilde{D}_{\rm home,end}=D_{\rm home,end}/D_{\rm home,work}$, and likewise for walking mid points. 

A major advantage of the TimeUse+ survey is the active diary component where participants record information including location types (home, work, or other), trip purposes, expenditures, and partners of each trip and location. Participants are also asked to detail the activities (e.g., exercising, shopping, eating) and their durations at each location. This largely enriches the dataset compared to other studies which solely rely on GPS. The full variables of the survey are described at  \cite{winkler_timeuse_2023}. Together with the detailed information on participants' background (e.g., employment status, access to car, public transport subscription, having kids, having dogs, views on green transport), the TimeUse+ dataset allows to analyze the association between personal features and walking. The survey collects in total 225 variables on the participants through questionnaire and on each activities. To avoid possible detection error and outliers, we remove the top 0.5 \% trips in terms of non-stop walking distance (above 16.8 km) or duration (above 101 minutes) and remove those with instataneous speed above 9.6 km/hr \cite{rodriguez_identifying_2012}. 

To quantify the travel flows in each area, we calculate gridded track density as the average GPS track crossing from all participants, as described in the following:
      1. Create grids for the considered domain: 500m resolution for entire Switzerland and 50m resolution for the city of Basel-Stadt
      2. Rasterize participants' tracks by counting crossings per grid. The grid accumulates 1 per track crossing
      3. Divide the counts of each participant by number of days they ever walk in the domain
      4. Apply Gaussian smoothing to account for GPS tracking uncertainty
      5. Take the average over the crossing counts from participants 
Track density is calculated for different travel modes: walking, cycling, and driving. The track density raster is then compared with layers of population and building density. This method outputs a clear visualization of massive GPS tracks and allows flexibility to adjust spatial resolution. 

\subsection{Dual-center EPR model}
To describe the observed mobility patterns, we model an individual’s walking behaviour as a sequence of visitation events
$\{(t_i,\mathbf{x}_i)\}_{i=0}^{N-1}$, where $N$ is the total number of events in the simulated trajectory, $\mathbf{x}_i\in\mathbb{R}^2$ is the location visited at event $i$ (end of a trip / start of an activity episode), and $t_i$ is the corresponding event time. The proposed model extends the Exploration and Preferential Return (EPR) mechanism \cite{song_modelling_2010} in three ways that are motivated by the TimeUse+ observations: (i) two dominant anchors representing home and work, (ii) a simple circadian schedule that increases the probability of being at the appropriate anchor at the appropriate time, and (iii) state-dependent waiting times to reproduce empirical time allocation at home, work, and other places.

The model keeps track of the set of distinct locations visited up to event $i$,
$\mathcal{P}_i=\{\mathbf{x}^{(1)},\dots,\mathbf{x}^{(S_i)}\}$, where $S_i=|\mathcal{P}_i|$ is the number of unique locations visited up to that event and $\mathbf{x}^{(j)}$ denotes the coordinates of the $j$th distinct location. Thus, $\mathbf{x}_i$ denotes the location occupied at event $i$, whereas $\mathbf{x}^{(j)}$ identifies one of the distinct locations stored in $\mathcal{P}_i$. The model also keeps track of the corresponding visit counts $\{c_j(i)\}_{j=1}^{S_i}$, where $c_j(i)$ is the number of visits to location $\mathbf{x}^{(j)}$ up to event $i$. We initialize the set of locations with two anchors, $\mathbf{x}_{\mathrm{home}}$ and $\mathbf{x}_{\mathrm{work}}$, and assign them large initial counts so that they act as highly recurrent locations from the beginning; all newly discovered locations are added with initial count $c=1$. 

Time is introduced through inter-event waiting times. Between two consecutive events the agent remains at the current location for a duration $\Delta t_i=t_i-t_{i-1}$, which we interpret as the dwell time at the state occupied at event $i-1$. We sample $\Delta t_i$ conditional on the current state $s_{i-1}\in\{\text{home},\text{work},\text{other}\}$:
\begin{equation}
\Delta t_i \sim \psi_{s_{i-1}}(\cdot),
\end{equation}
where $\psi_s$ denotes the waiting-time probability distribution associated with state $s$. In the empirical implementation, $\psi_s$ is estimated from activity durations reported in TimeUse+ by location type (home/work/other). It can also be specified parametrically (e.g., exponential \cite{wang_extended_2019} or lognormal distribution \cite{jin_urban-epr_2025}) when an empirical distribution is not available. 

The circadian routine is implemented by evaluating the hour-of-day of the updated event time $t_i$ and defining an indicator $I_{\mathrm{work}}(t_i)$ that equals 1 during predefined work hours (8 am-5 pm) and 0 otherwise. With probability $\sigma\in[0,1]$ (schedule strength), the next location is forced to the corresponding anchor:
\begin{equation}
\mathbf{x}_i=
\begin{cases}
\mathbf{x}_{\mathrm{work}}, & \text{with prob. }\sigma \text{ if } I_{\mathrm{work}}(t_i)=1,\\
\mathbf{x}_{\mathrm{home}}, & \text{with prob. }\sigma \text{ if } I_{\mathrm{work}}(t_i)=0,
\end{cases}
\end{equation}
which encodes the dominant home--work rhythm while allowing stochastic deviations controlled by $\sigma$. With the complementary probability $1-\sigma$, no schedule-induced snap occurs and the next location is generated by the standard EPR mechanism \cite{song_modelling_2010}. We set $\mathbf{x_{home}}$ at (0,0) and $\mathbf{x_{work}}$ at (1,0). At event $i$, the probability of exploring a new location decreases with the number of distinct previously visited locations:
\begin{equation}
p_{\mathrm{new}}(i)=\rho\, S_{i-1}^{-\gamma},
\end{equation}
where $\rho\in(0,1)$ controls the baseline exploration rate and $\gamma>0$ controls how fast exploration
saturates as the repertoire of familiar places grows. 

If exploration occurs, with probability $p_{\mathrm{new}}(i)$, a new location is generated by a random step from the previous position:
\begin{equation}
\mathbf{x}_i=\mathbf{x}_{i-1}+\Delta\mathbf{x},\qquad
p(\|\Delta\mathbf{x}\|) \propto x^{-\alpha} \, e^{-\lambda x}, \qquad x \in [x_{\min}, x_{\max}] ,
\end{equation}
with a uniformly distributed direction and a truncated power-law jump length $\|\Delta\mathbf{x}\|$, characterized by the power-law exponent $\alpha$ and exponential cutoff parameter $\lambda$, and bounded by $[x_{\min},x_{\max}]$. We fit these parameters using TimeUse+ tracks. First, we calculate the distance between each trip endpoint and the previous trip endpoint as $\|\Delta\mathbf{x}\|$. Second, we normalize this distance by the home-work distance $D_{\rm home,work}$, as described previously, to obtain the normalized jump length $\Delta\tilde{x}=\|\Delta\mathbf{x}\|/D_{\rm home,work}$. Third, we use the Python module \texttt{powerlaw} \cite{alstott_powerlaw_2014} to fit the distribution and estimate $\alpha$ and $\lambda$. Finally, the fitted parameters are used as inputs to the EPR simulation. This choice produces frequent short displacements and occasional longer ones while respecting the empirical support of walking distances.

If exploration does not occur, the agent returns to a previously visited location $\mathbf{x}^{(j)}\in\mathcal{P}_{i-1}$ via preferential return, with probability proportional to its past visit frequency:
\begin{equation}
\Pr\left(\mathbf{x}_i=\mathbf{x}^{(j)}\right)
\propto \big(c_j(i-1)\big)^{\alpha},
\end{equation}
where $\alpha>0$ (return exponent) controls how strongly the process concentrates on habitual locations.

To reflect the empirical fact that returns are not time-homogeneous, we optionally apply a mild time-of-day
bias by upweighting the work anchor during work hours and the home anchor outside work hours before
normalizing the return probabilities. The simulated output is a synthetic sequence of locations and times from which we compute walking-relevant diagnostics: jump-length distributions (walking displacements), walking probability maps, time-weighted spatial heatmaps (time spent around home/work versus elsewhere), and inter-event waiting-time distributions by states. We also calculate Shannon entropy, which measures the unpredictability of spatial distribution by $H = -\sum_{i=1}^{n} P(\mathbf{x_i}) \log P(\mathbf{x_i})$, where $\mathbf{x_i}$ is each unique location and $P$ is the probability of visiting there. The full parameter set is $\{\rho,\gamma,\alpha,\lambda,x_{\min},x_{\max},\sigma,
\psi_{\text{home}},\psi_{\text{work}},\psi_{\text{other}},\text{work hours}\}$. The model is evaluated with spatial distribution of walking trips with the TimeUse+ dataset.  

\subsection{Data on environmental features}
To analyze the effects of urban characteristics on walking behavior, we also compile information on residential population density, public transport, and walkable road intersection. All these features have previously been shown to correlate with walking behaviour \cite{cervero_travel_1997, cerin_determining_2022} and are obtained from the following official sources. Population density is based on the STATPOP residential population at 100-m resolution \cite{noauthor_population_2021}. Counts of public transport stops are calculated from SwissTLM3D, a dataset containing all public transport stops in Switzerland \cite{noauthor_swisstlm3d_2024}. Counts of road intersections are calculated from the road network layer from the same dataset, excluding highways (not walkable). The residential environmental feature is calculated for 1-km buffer around the home location of each participants. Apart from density, we use the Swiss municipal classification of urban-rural typology \cite{federal_statistics_office_typology_nodate}, which uses both morphological (population density and size) and functional (e.g., commuter flows and accessibility) features to classify municipalities into 3 classes and 9 sub-categories: urban area (large, medium, and small) , periurban area (high, medium, and low density), and rural area (central, local center, and remote). Each walking location reported in TimeUse+ and participants' home are assigned a typology based on their main municipality class. 

\section{Results}
\subsection{Spatial distribution of walking}
\begin{figure}[h]
\centering
\includegraphics[width=1\textwidth]{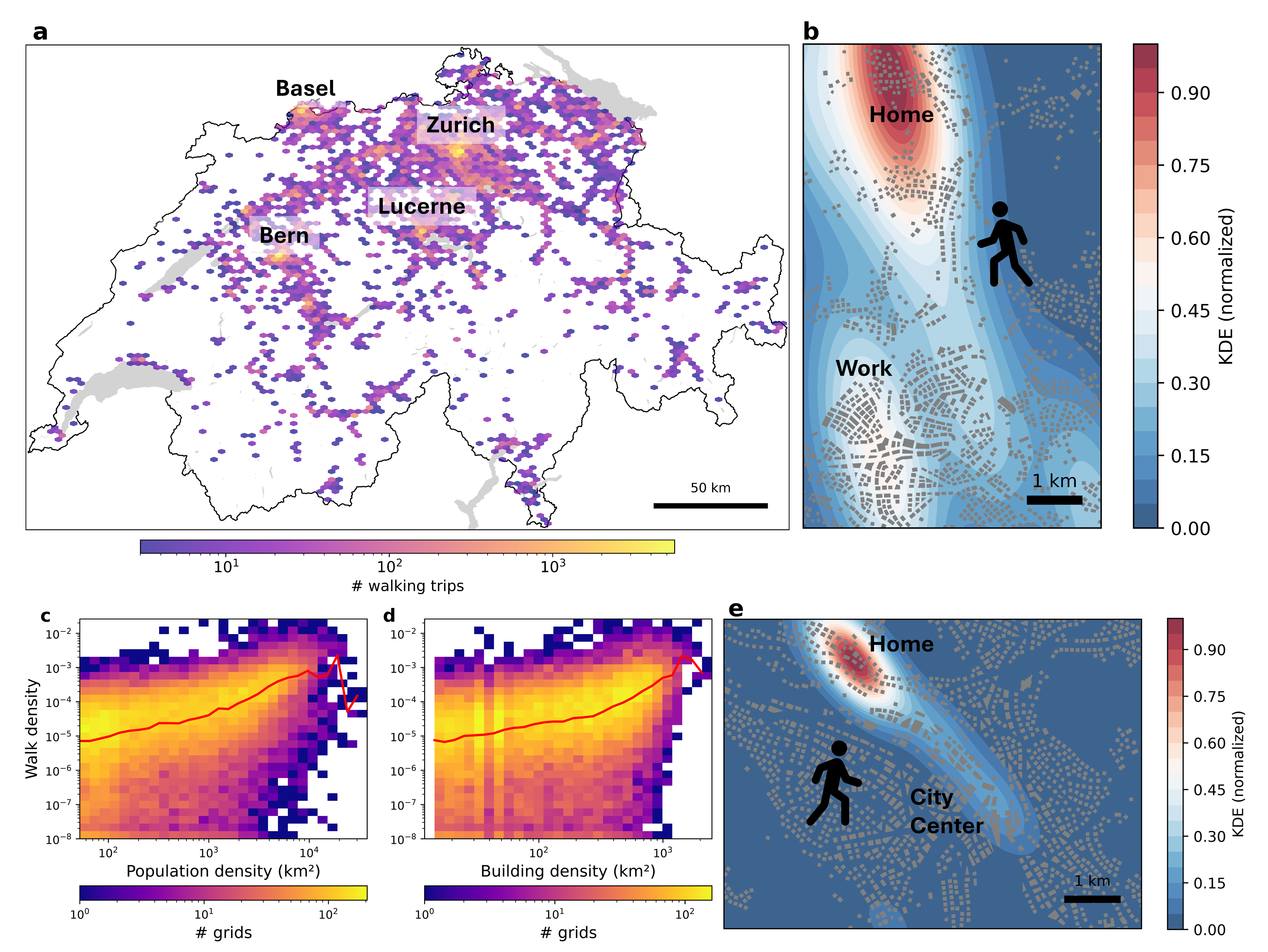}
\caption{ {\bf Spatial distributions of walking trips.} (a) All walking mid points from the TimeUse+ (b) Normalized KDE of walking mid points of an anonymized employed participant (c \& d) Relationships between population/building density and walk density at 500m resolution across Switzerland. Red curves show the median walk density across the 500m grids within the population/building density bins (quantile binning) (e) Normalized KDE of walking mid points of an anonymized retired participant
}\label{fig1}
\end{figure}

Walking is the most common mobility mode in terms of trip number and the second most common (after car) in terms of total duration in the TimeUse+ survey \cite{winkler_timeuse_2024}. Fig.~\ref{fig1}a shows the heat map of walking trips from TimeUse+. As most participants live in or near the urban areas of Zurich, Basel, Lucerne, and Bern, these areas record the highest density of walking trips while regions further away, such as the Southern and Eastern Switzerland, have the lowest event density. Zooming into one profile of an employed person, Fig.~\ref{fig1}b shows the typical distribution patterns of walking trips for someone with a fixed residential location and workplace. The majority of walks are close to the home locations, followed by some near the workplace, and the rest of the walking trips are scattered in other less-visited locations. Another profile of a retired person (Fig.~\ref{fig1}e)  shows high walking density near home and, replacing workplace, the city center. More walking near home, commonly-visited locations like workplaces, and city centers are typical behaviours also found in several previous studies (e.g., \cite{millward_active-transport_2013}). The data also shows that most walks occur in areas of high population and building density, which are generally richer in amenities and services (restaurants, shops, offices, etc.) (Fig.~\ref{fig1}c\&d). The nonlinear relationships between density and walking and its reduction at the densest areas aligns with previous literature \cite{shen_systematic_2025, gao_what_2020, saelens_built_2008, agrawal_extent_2007}.

\begin{figure}[h]
\centering
\includegraphics[width=1.0\textwidth]{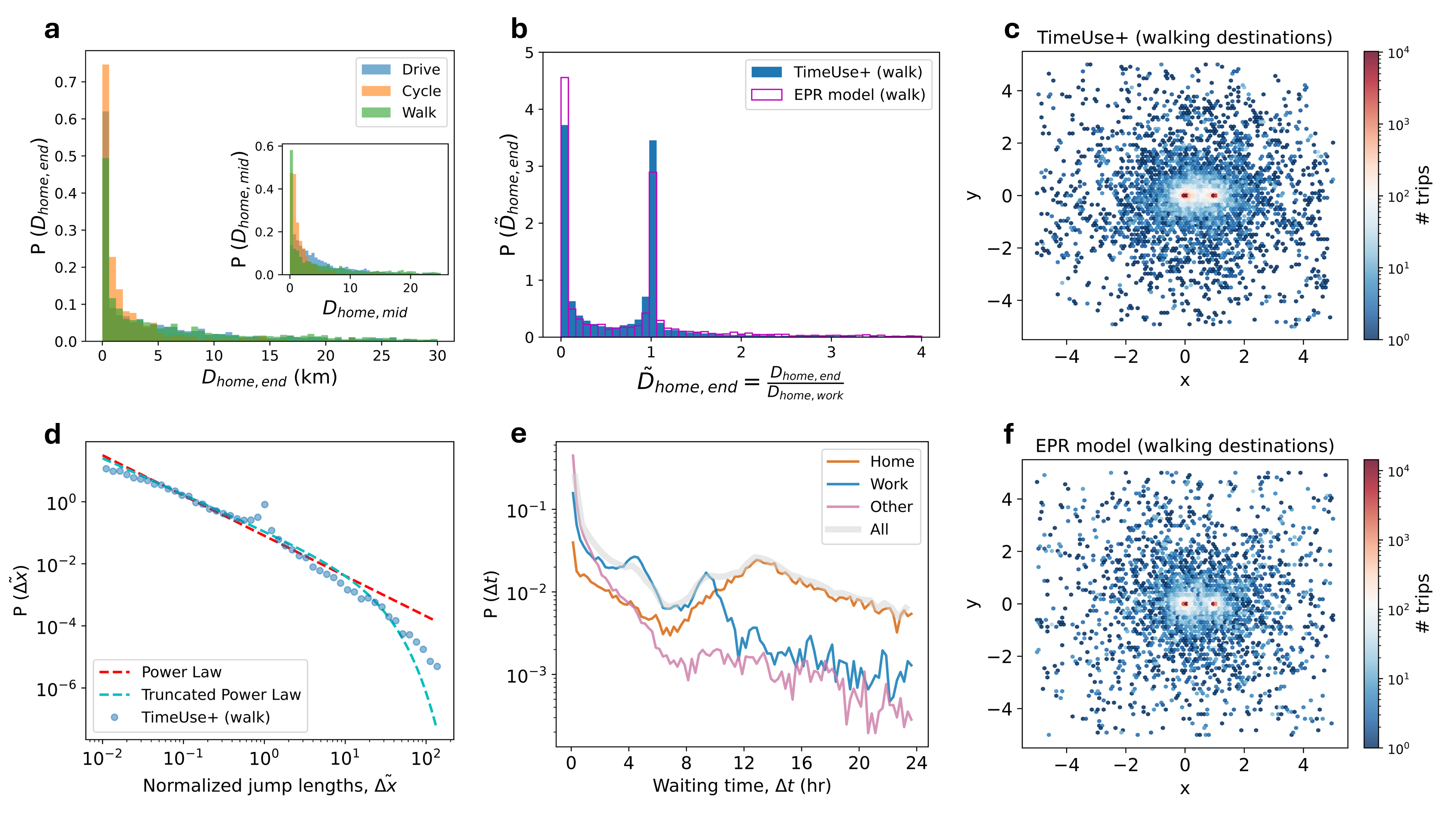}
\caption{{\bf Probability distribution of walking.} (a) PDF of the distance between walking end or mid points to home ($D_{\rm home}$) (b) PDF of the normalized distance between walking endpoint and home ($\tilde{D}_{\rm home,end}$) (c) spatial distribution of TimeUse+ walking end points after normalization (d) PDF of normalized jump length ($\Delta \tilde{x}$) between consecutive walking end points (e) empirical waiting time distribution from TimeUse+ (f) spatial distribution of walking destinations in the dual-center EPR model. The results in (b-d) excludes weekends, unemployed participants, and whose home and workplace are closer than 1 km, leading to 671 participants.}\label{fig2}
\end{figure}

Overall, most walks are very close to individuals' home - with 35.1 \% of walking end points (34.6 \% of midpoints) within 1 km from home and 55.3 \% within 5 km (Fig.~\ref{fig2}a). The rapid decay of trip numbers with the distance from home ($D_{\rm home}$) also holds for other transport modes, as all modes are commonly used to reach home. However, solely using the decay relationship from Fig~\ref{fig2}a (as commonly done in literature e.g. \cite{iacono_access_2008}) may overlook other common walking hot spots that are away from home. As a matter of fact, after normalizing $D_{\rm home,end}$  with the distance between home and workplace (i.e., $\tilde{D}_{\rm home,end}=D_{\rm home,end}/D_{\rm home,work}$ for each employed participant whose home and workplace are at least 1 km apart), we found that another area of high-density walking endpoints emerges near workplaces ($\tilde{D}_{\rm home,end}=1$) (Fig.~\ref{fig2}b). After this normalization, the probability distribution of walking locations of the 671 employed participants collapse to a bimodal (or dual-center) distribution, showing one distance-decay relationship from home and another one from workplace (Fig~\ref{fig2}b). Most walks happen within the range of  $0\leq\tilde{D}_{\rm home,end}\leq1$ while walks further away are rare. 

Normalizing by $D_{\rm home,work}$ and anchoring home and workplace locations at (0,0) and (0,1) on the (x,y) plane, we can then linearly transform every employed participants' GPS-recorded walking midpoints, showing that the 2-D probability distribution collapses onto two well-defined walking hotspots: home and workplace (Fig.~\ref{fig2}c). Most walks are located near the two anchors and mostly along the home-workplace axis, similar to previous studies on overall mobility \cite{gonzalez_understanding_2008, wang_extended_2019}. The walks near workplaces decays faster with distance compared to home, especially at places above $\tilde{D}_{\rm home,end}=1$. Note that the color bar is in log scale, meaning that walks further away from the anchors are quite rare. 

\subsection{Fitting and evaluating the dual-center EPR model}
Based on this finding, we develop the dual-center EPR model described in the Methods section. We start with analyzing the distance between two consecutive walking destinations, which we defined as jump length $\|\Delta\mathbf{x}\|$. Fig.~\ref{fig2}d shows that the empirical normalized jump lengths ($\Delta \tilde{x}$) from TimeUse+ follows a truncated power law distribution with  $\alpha=1.16$  and $\lambda=0.06$ (except a peak near workplace where $\Delta \tilde{x}=1$). This means that most jumps are short with occasionally long jumps. The empirical waiting time at different locations follow clearly different distributions (Fig.~\ref{fig2}e). While waiting time at 'other' location may be a power-law distribution, waiting times at home (peaking at 14 hours) and workplace (peaking at 4 and 9 hours) have regular weekday diurnal patterns. The separation of location types thus allows for more realistic walk event generation in time that otherwise would not be captured (gray curve in Fig.~\ref{fig2}e). The fitted $\alpha$ and $\lambda$ and the empirical state-dependent waiting time distribution are input to the EPR model. 

We evaluate the EPR simulation results by comparing $\tilde{D}_{\rm home,end}$ (Fig.~\ref{fig2}b) and the frequency heatmap of walking locations (Fig.~\ref{fig2}f) with the TimeUse+. Fig.~\ref{fig2}b shows that the proposed model captures well the empirical distributions of walking. The probability decays with distance further away from the two anchors. This is also shown by the heatmap, which demonstrates that the model can reproduce the 2-D distribution pattern of  walking hotspots (Fig.~\ref{fig2}f). Most walks are near the anchors and along the axis between them. We also plot the spatial distribution of waiting time, which are similar to Fig.~\ref{fig2}f but with even higher values at home and work as most time are spent there (Fig.~SI\ref{sfig1}). To show the improvements from inputting empirical waiting times, the modelled waiting time distribution at each location types with and without inputting empirical data is shown in Fig.~SI\ref{sfig2}.

\subsection{Spatial distribution of different mobility modes}
\begin{figure}[H]
\centering
\includegraphics[width=1\textwidth]{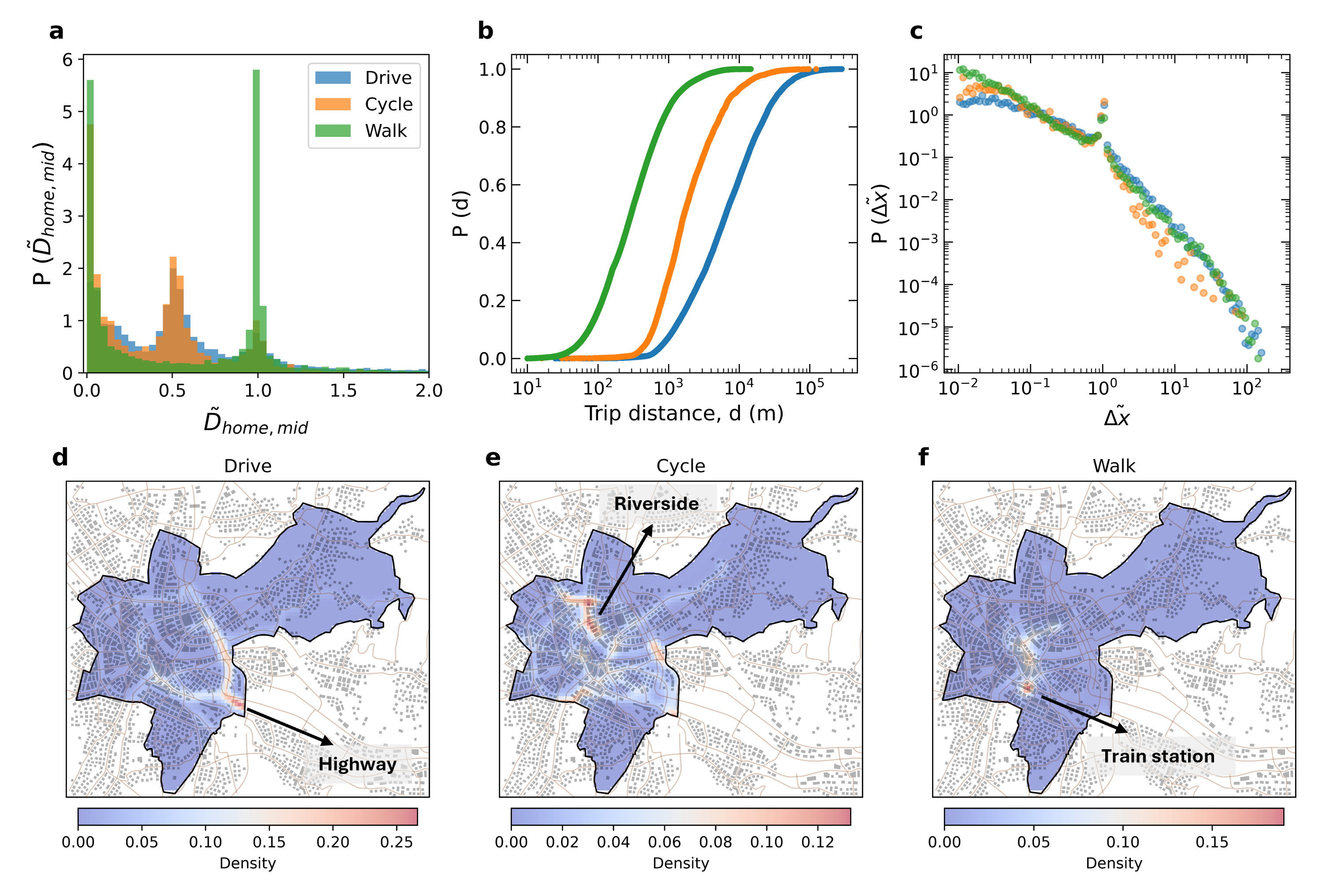}
\caption{{\bf Spatial distribution of different transport modes.} (a) PDF of normalized distances between trip midpoints to home for the three transport modes (b) CDF of trip distances for each transport mode (c) PDF of normalized jump lengths for each mode. (d) - (f) Heatmaps showing the counts of trip endpoints of each mode in Basel, Switzerland (trips from participants living in Basel are excluded).}\label{fig3}
\end{figure}

The consideration of the normalized walking-point-to-home distance $\tilde{D}_{\rm home}$ provides a new lens to inspect the spatial distribution of each transport mode. While Fig~\ref{fig2}a seems to suggest similar decay relationships with $D_{\rm home}$ for different transport modes, their distinctions unfold after normalization (Fig~\ref{fig3}a). The high-density of trajectory midpoints of cycling and driving is concentrated near the middle between home and workplace ($\tilde{D}_{\rm home,mid}$=0.5). Thus, these transport modes cover almost the whole commuting distance, emphasizing their role as main transport means connecting the two walking hotspots. Whereas for walking, the distribution closer to home and workplace suggests that it is used as the first-/last-mile of the home-work trips or just short trips near these two major mobility anchors. Note that there are also short driving or cycling trips just near the home location but it is much less pronounced than the peaks near $\tilde D_{\rm home,mid}$=0.5 (Fig~\ref{fig3}a). Unlike the clear dual-center structure of walks, driving and cycling midpoints show only weak peaks around the workplace. This can be explained by the much shorter trip distance of walking compared to cycling and driving Fig~\ref{fig3}b. Walking trips have a median distance of 293 meters and dominate trips below 400 meters while the median distances of cycling and driving trips are 1701 and 6586 meters, respectively. Table.~SI.\ref{tab:event_distance} shows the complete statistics for each mode. Comparing the $\Delta \tilde{x}$ for each mode (Fig.~\ref{fig3}c), cycling and driving show lower probability at short distance to home, while all modes show peaks near 1 (home-work distance). At $\Delta \tilde{x}>1$, walking and driving show similar distributions but cycling is less common.    

Using the city of Basel-Stadt (37 $km^2$) as an example, we show how the use of different transport modes is spatially distributed in a real city. We plot the distribution of all TimeUse+ mobility tracks within the boundary of Basel after excluding the 25 participants living there to avoid bias towards where they live (Fig.~\ref{fig3}d-f). The highest density of driving is near the highway entry/exits and its connecting roads, where walking density is low. Cycling shows more dispersed distribution from both the central transit and commercial areas, showing its more diverse usage and longer travel distance. Particularly, higher cycling density is observed near riverside and bridges probably because these places appeal more to cyclists. The highest density of walking appears near the Basel main train station, followed by the central business district to the North, which aligns with the findings in Fig~\ref{fig1}.  These diverging patterns clearly demonstrate that the regularity observed in Fig~\ref{fig2} emerge from significant intra-urban variability, where mobility patterns are shaped by the usage, purpose, and spatial distribution of points of interest, infrastructures, and transport modes. 

\subsection{Effects of personal and environmental features on walking} 
\begin{figure}[H]
\centering
\includegraphics[width=1\textwidth]{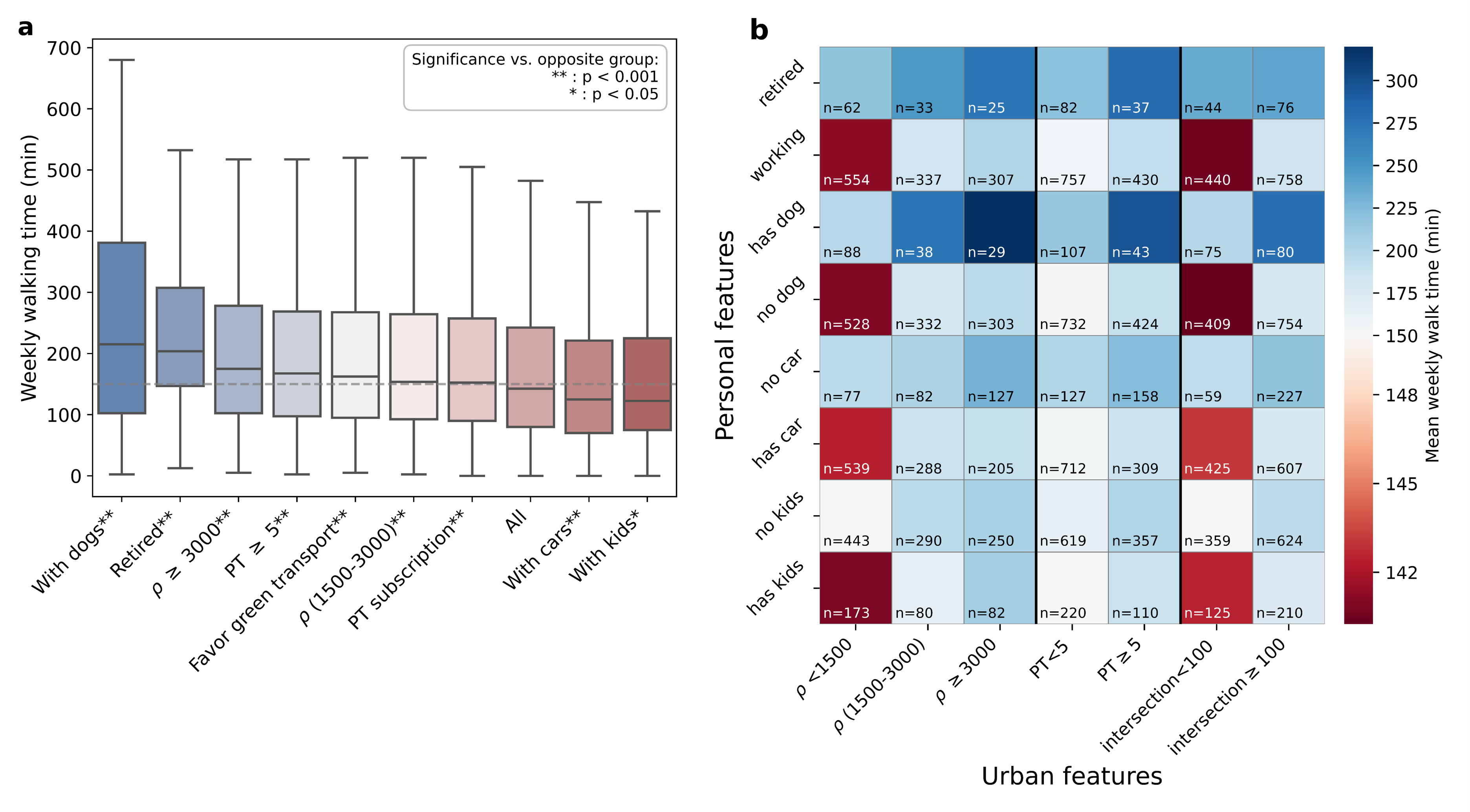}
\caption{{\bf Walking durations of different groups classified by personal and residential environmental features.} (a) Box plots of weekly walking time for participants in different groups. (b) Heatmap of mean weekly walking time for groups combining of different personal and density-related environmental features}\label{fig4}
\end{figure}

We now turn our attention to the different personal and environmental features that influence walking behaviour. In Fig.\ref{fig4}a we show the box plot of walking times for population subgroups with different personal features and residential environments. In terms of personal features, the subgroups with a car or children (under 12 years) have a shorter walking time, while those who have a dog or have retired walk more. Owning public transport subscriptions increases walking time, showing the positive effect of public transit for walking. Aligning with this result, people who favor green transport (i.e., those who answered 'extremely important' or 'very important' to 'changing individual transportation behavior to mitigate climate change' in the TimeUse+ questionnaire \cite{winkler_timeuse_2024}) also walk more than the average. In fact, about 30\% of walking activities during the TimeUse+ survey took place right before or after using public transport, thus proving the strong inter-dependence between walking and transits. One perhaps unexpected finding different from some previous studies (e.g., \cite{chan_daily-life_2024, adjaye-gbewonyo_walking_2024}) is that the retired population walk much more than the average. This result can be explained by the fact that seniors in high-income countries have more free time to walk and, in the case of this study, they are not too old to experience difficulties in walking (only 15 \% retired participants in TimeUse+ are above 75 years old) and can thus keep a physically active life style.

As for environmental features, density of population and public transit stops in the residential neighborhoods of the participants has positive association with their total walking time. Living in denser neighborhoods ($>$ 1500 persons $/km^2$, an international criteria for urban centers \cite{eu_applying_2021}) increases walking time. People living in areas of even higher density ($>$ 3000 persons $/km^2$) walk even more. Living in areas of high public transport density (above 5 stops per km$^2$) also increases walking time, which aligns with previous analysis on the positive effect of walking for transits. These observations confirm that, on average, people living in dense urban areas walk more than people living in rural area, which is consistent with previous studies (e.g., \cite{cerin_determining_2022}). Note that the association here is between walking duration of each participant and their home neighborhoods, which is not the same as the relationships between walking density (calculated from walking tracks of all participants) and population/building density shown in Fig.~\ref{fig1}c\&d. 

To understand the joint effects of personal and environmental features, we further analyze weekly walking time of different population groups (Fig~\ref{fig4}b). Results show that higher densities of population, public transport stops, and road intersections increase walking time for all groups - regardless of their different personal features considered here. At the same time, the influence of personal features on walking time generates the same outcomes regardless of the characteristics of residential environments. For example, people having a car and living in areas with the same density, always walk, on average, less than people without a car. This implies that although urban design can prompt active mobility, behavioral and societal changes (e.g., lower car dependency) are still needed to achieve healthy walking time. The groups that walk the least, and do not achieve the threshold of 150 minutes per week, are those with both personal and environmental features negatively correlated with walking (e.g., having young kids or cars and living in areas of low density). Such groups should be the target of urban planning policies aimed at increasing physical activity. Nevertheless, in some circumstances, living in denser urban areas may offset the walking time loss due to certain personal features. For example, we find that parents with children living in dense neighborhoods ($>$ 3000 persons $/km^2$) walk 208 minutes per week, even slightly more than people without kids living in the same urban environments. Therefore, dense urban areas - and perhaps higher densities of family-friendly facilities - may encourage more walking for parents with young children.  

\begin{figure}[!h]
\centering
\includegraphics[width=1\textwidth]{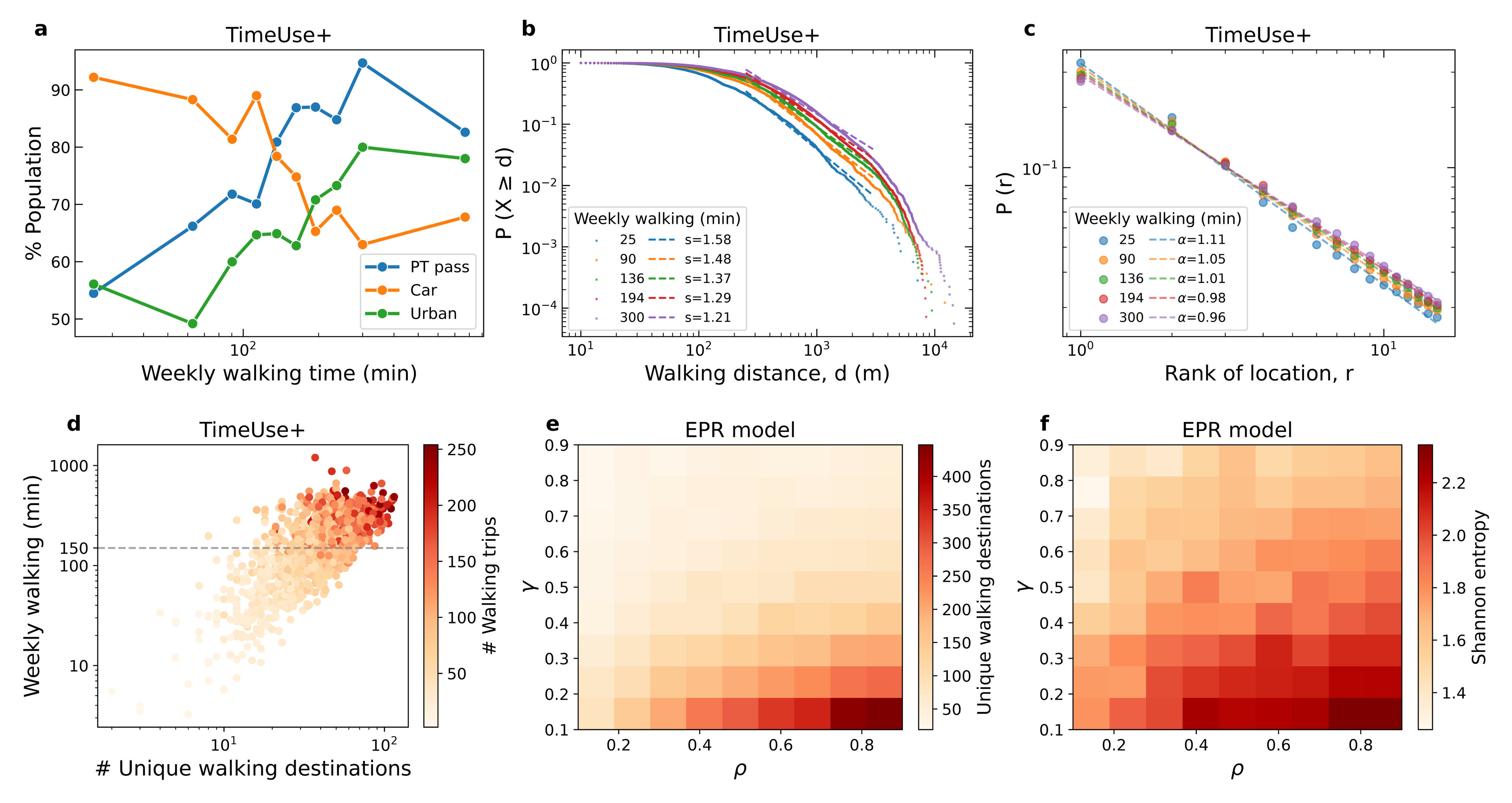}
\caption{{\bf Variations in walking behaviors in TimeUse+ and the EPR model.} (a) Demographic transitions across ten walking-time deciles. (b) CCDF of walking distances and the fitted slopes (s) (fitting range: 250 $<$ x $<$ 3000 m). (c) Frequency-rank relationships of unique trip destinations and the fitted exponents. (d) Association between the number of unique walking destinations and walking time. (e), (f) The number of unique walking locations and Shannon entropy simulated by the EPR with varying input parameters} \label{fig5}
\end{figure}

We group the participants into 10 deciles based on weekly walking time, which confirms the shift in travel behaviour and living environments. The groups that walk more show a much higher percentage of public transit pass subscription, lower car ownership, and mostly living in urban environments (Fig.~\ref{fig5}a). Among those who walk above 120 minutes weekly, higher ratio of population have public transport pass than car. The total walking time is also reflected in the walking distance of each trip, as illustrated by the complementary cumulative distribution functions (CCDF) for each group (Fig.~\ref{fig5}b). Groups with higher total walking times show a rightward shift in their distribution functions, indicating a higher composition of longer-distance walks, whereas less-walking groups are primarily characterized by shorter walks. In other words, those who "accept" longer walking distance per trip show a higher total walking time. Particularly, the median trip distance increases nearly threefold from the lowest walking group (145 m) to the highest (424 m) (see CDF in Fig.~SI\ref{sfig4}a). This is also revealed by the fitted slopes to the CCDF for each decile, which increase as weekly walking time reduces, showing their lower probability to take longer walk. 

\subsection{Walking destinations, exploration, and preferential return} 
Another factor relating with the total walking time is whether one's mobility spreads across numerous unique locations (i.e., explorer) or concentrate at a few places (i.e., returner). This can be illustrated by the frequency-rank probability distribution: $P(r)\sim r^{(-\alpha)}$, where $r$ is the rank of the location based on the frequency of visits and $\alpha$ is the fitted exponent \cite{gonzalez_understanding_2008, song_modelling_2010}. Fig.~\ref{fig5}c shows the visit probability to the top 15 locations ranked by frequency for each walking group in TimeUse+. The groups that walk less have a higher value of $\alpha$, indicating that their travels are mostly at highly-frequented locations and only take a few trips to other places. On the other hand, the groups with higher walking time show lower $\alpha$, meaning more dispersed travel patterns and higher probability to visit less-frequented places. Aligning with this finding, Fig.~\ref{fig5}d shows that the TimeUse+ participants with a higher number of unique walking destinations (and total number of walking trips) have longer total walking time. This suggests that people who explore more destinations on foot have higher walking time compared to people who only walk at a few locations (or barely walk). 

Such behavioral differences can be modelled using the proposed dual-center EPR modeling framework, in which the number of unique visited locations are largely controlled by the modelling parameters $\rho$ (overall explore tendency) and $\gamma$ (how fast the exploration decreases as known places increase). Fig.~\ref{fig5}e shows that, as expected, higher $\rho$ and lower $\gamma$ lead to higher number of unique visited places, which is empirically associated with higher walking time (Fig.~\ref{fig5}c). The exploration tendency is also captured by the Shannon entropy, which quantifies the unpredictability of spatial distribution (See Methods). In the EPR model, Shannon entropy shows the same trend as the number of unique locations, increasing with higher $\rho$ and lower $\gamma$  (Fig.~\ref{fig5}f). Empirically, also the TimeUse+ participants with higher Shannon entropy of walking destinations tend to have higher total walking time (Fig.~SI. \ref{sfig4}b). Overall, these modelling and empirical results imply that 'explorers' who tend to visit more places, especially those exploring more by foot, have higher walking time than 'returners' who restrict their mobility and walking to a few locations. 

\section{Discussions}
Using mode-separated and location-annotated GPS tracks, this study shows a dual-center distribution of walking locations, which can be reproduced by an EPR model incorporating a circadian schedule and two anchors: home and workplace (Fig.~\ref{fig2}). Furthermore, thanks to the high-resolution of the GPS data, we reveal the existence of distinct mobility hot spots, which vary depending on different transport modes (Fig.~\ref{fig3}). Such information can hardly be captured by other (coarser) mobility data sources. 
The observed bimodal distribution of walking activities have important implications for urban studies, from modelling residents' mobility \cite{lin_modelling_2026} to outdoor heat and air pollution exposure \cite{lin_daily_2025,nyhan_exposure_2016}. For example, our EPR framework can be applied to agent-based models to improve the simulation of walking behaviour, which is often simplified in mobility models (e.g., \cite{lin_modelling_2026}). By analyzing and simulating the distribution of walking activities, this work can also guide the development of policies to improve cities' walkability - for example, targeted interventions along the axis between the main residential and business districts may be an effective way to increase walking ratio. In addition, as people who accept longer walking distance and with more 'explorer' characters have higher average walking time (Fig.~\ref{fig5}b,c,d), prompting population to explore nearby/new locations (e.g., by enhancing public transport convenience and diversifying land use and points of interest in proximity) may contribute to increase active mobility and health. In this regard, considering the full walking paths or, at least, the main walking hot spots (Fig.~\ref{fig2} and~\ref{fig3}), can reduce biases related to the "residential effect fallacy" \cite{chaix_residential_2017} and advance the identification and mitigation of climate-related risks, such as extreme heat and air pollution \cite{yin_dtex_2021, park_individual_2017, reis_influence_2018}. 

In general, our study reveals the synergistic effects of personal and environmental features on walking time. Residents in dense urban areas walk more than people living in rural areas (Fig.~\ref{fig4}) (as previously found, e.g., \cite{cerin_determining_2022, frank_linking_2005}) but personal features also have strong effects on walking time (aligning with e.g., \cite{brown_relationships_2006,sehatzadeh_walking_2011}). Our cross evaluation shows that both urban design and behavioral shift are needed to achieve a physically active life style and different strategies should be adopted for different population groups. For those having a car, living in denser areas can increase their walking by reducing car dependency. The positive association between public transport subscription and walking time suggest that walking for transit can contribute much to the gap in walking time, echoing with suggestions from previous studies \cite{lachapelle_transit_2009,freeland_walking_2013}. Therefore, intervention like adjusting public transport price and adding stops may be able to encourage walking. As personal value towards green transport is also associated with increased walking time, education and campaigns that promotes its use may also lead to more active mobility. In fact, urban areas function as popular walking destinations even for people not living there. Our analysis on inter-city walking destinations shows that those living in rural and suburban areas walk often in urban areas (Fig.~ SI\ref{sfig3}) where urban residents are also walking the most. This implies that although most walks are near residential neighborhoods, urban areas attract disproportionately more walking. Putting this in the context of the dual-center distribution, this suggests that urban areas attract more walking because they are hotspots of workplaces. Thus, considering only residential neighborhoods for improving walkability may not be sufficient \cite{millward_active-transport_2013}. 

While informative, these results should be interpreted with caution due to the limitations of the present study. First, the sample size of 1,318 participants is smaller than other mobility studies using CDR, mobile Apps, or social media (but larger than other GPS-tracking studies e.g., \cite{rhee_levy-walk_2011, morabia_air_2009}). The samples are skewed towards younger individuals (18 to 40), higher education, and living in a single household, who may have a mobility behavior that differs from the population average \cite{winkler_timeuse_2024}. Second, we simplified the density calculation by using 1-km circle buffer around home rather than the 1-km walkable buffer used by previous studies \cite{cerin_determining_2022}. This should not have caused major difference as we only use density to observe the trend, rather than to derive a rigorous density-walking association as in e.g., \cite{cerin_determining_2022}. Third, it should be noted that Swiss cities are considered some of the most walkable cities in the world \cite{bruno_universal_2024} and they are mostly medium-sized in terms of population and density, so we do not cover the whole spectrum of rural-urban density variations. Finally, although there are clear walking behavioral differences among population subgroups, large variance exist (Fig.~\ref{fig4}a), limiting the ability to clearly distinguish between personal and environmental effects on walking behavior. In this context, there are also walkability factors beyond density that are not explicitly analyzed in this study, such as land use mix, design, and weather \cite{cervero_travel_1997, ewing_travel_2010, horanont_weather_2013}. 

\section{Conclusion}
Understanding the distribution of walking behavior and how personal and environmental features affect it is critical to improve healthy and low-carbon urban design. Using a GPS-tracked and diary-enriched time use survey dataset, we have identified a bimodal distribution of walking hot spots, with peak activities near home and workplaces. We have then developed a dual-center EPR model incorporating a circadian scheduler and waiting time sampler that is able to reproduce the observed walking patterns. We have also shown the intrinsically different distribution and characteristics of walking from other transport modes, which cannot be distinguished when using other types of mobility data. Overall, our findings suggest that solely improving walkability near residential areas may not be sufficient to achieve the health benefits of active mobility, as many walks occur near central business districts, where there are abundant job opportunities. Finally, we have diagnosed the synergistic effects of personal and residential features on walking time. This cross evaluation confirms that high-density urban environments attract most walks while individual features like car ownership, public transport subscription, having children or dogs, and being retired are also strongly associated with walking time. Therefore, urban design (e.g., densification and increasing walkable road connectivity), education (e.g., promote active and public transport), and life style changes (lower car dependency) are all feasible pathways to enhance walking. In addition, policy should prioritize the groups with both individual and environmental features negatively associated with walking, to help them reach healthy physical activity targets. 

\section{Acknowledgement}
The authors would like to thank Prof. Kay Axhausen and Daniel Heimgartner for providing the TimeUse+ dataset. We thank Quentin Gaillard on the methodology for calculating track density. GSL, GM, and MLLM acknowledge support from the SNSF Weave/Lead Agency funding scheme (grant number 213995).
\bibliographystyle{elsarticle-num} 
\bibliography{references}


\label{LastMainPage}
\clearpage
\pagestyle{empty}
\section*{Supplementary Material}
\setcounter{figure}{0}
\setcounter{table}{0}

\begin{figure}[!h]
\centering
\includegraphics[width=0.9\textwidth]{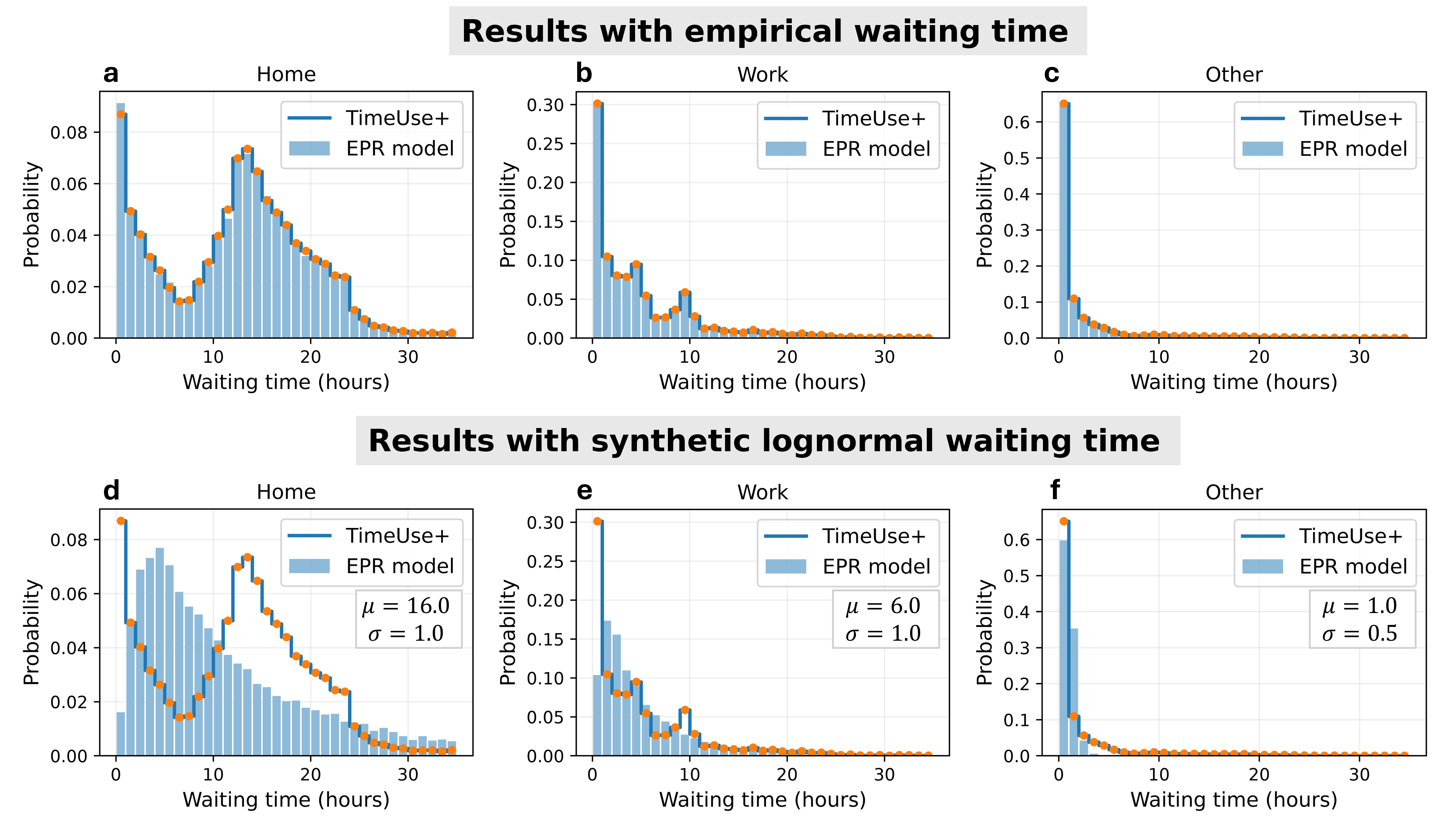}
\caption{ {\bf Probability distribution of waiting time} (a)-(c) probability distribution of waiting time from the EPR model when using TimeUse+ empirical state-dependent statistics. (d)-(f) probability distribution of waiting time from the EPR model when sampling from lognormal distribution with $\mu$  and $\sigma$  values in the figures. Empirical waiting time from TimeUse+ is shown in all figures for comparison.  
}\label{sfig1}
\end{figure}

\begin{figure}[!h]
\centering
\includegraphics[width=0.6\textwidth]{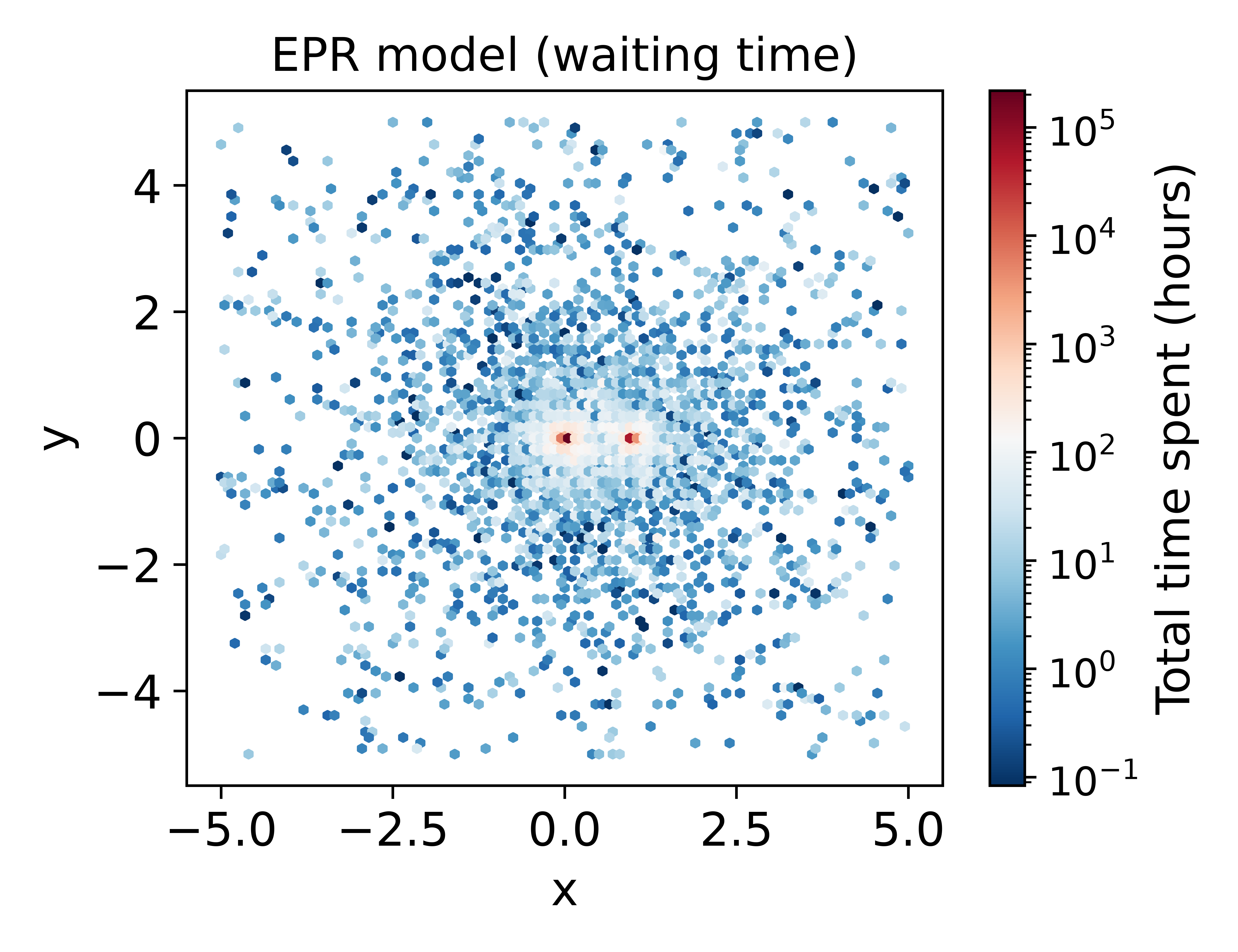}
\caption{ {\bf Heatmap of total waiting time from the EPR model.} Home location is at (0,0) and work location is at (1,0).
}\label{sfig2}
\end{figure}

\begin{figure}[!h]
\centering
\includegraphics[width=0.9\textwidth]{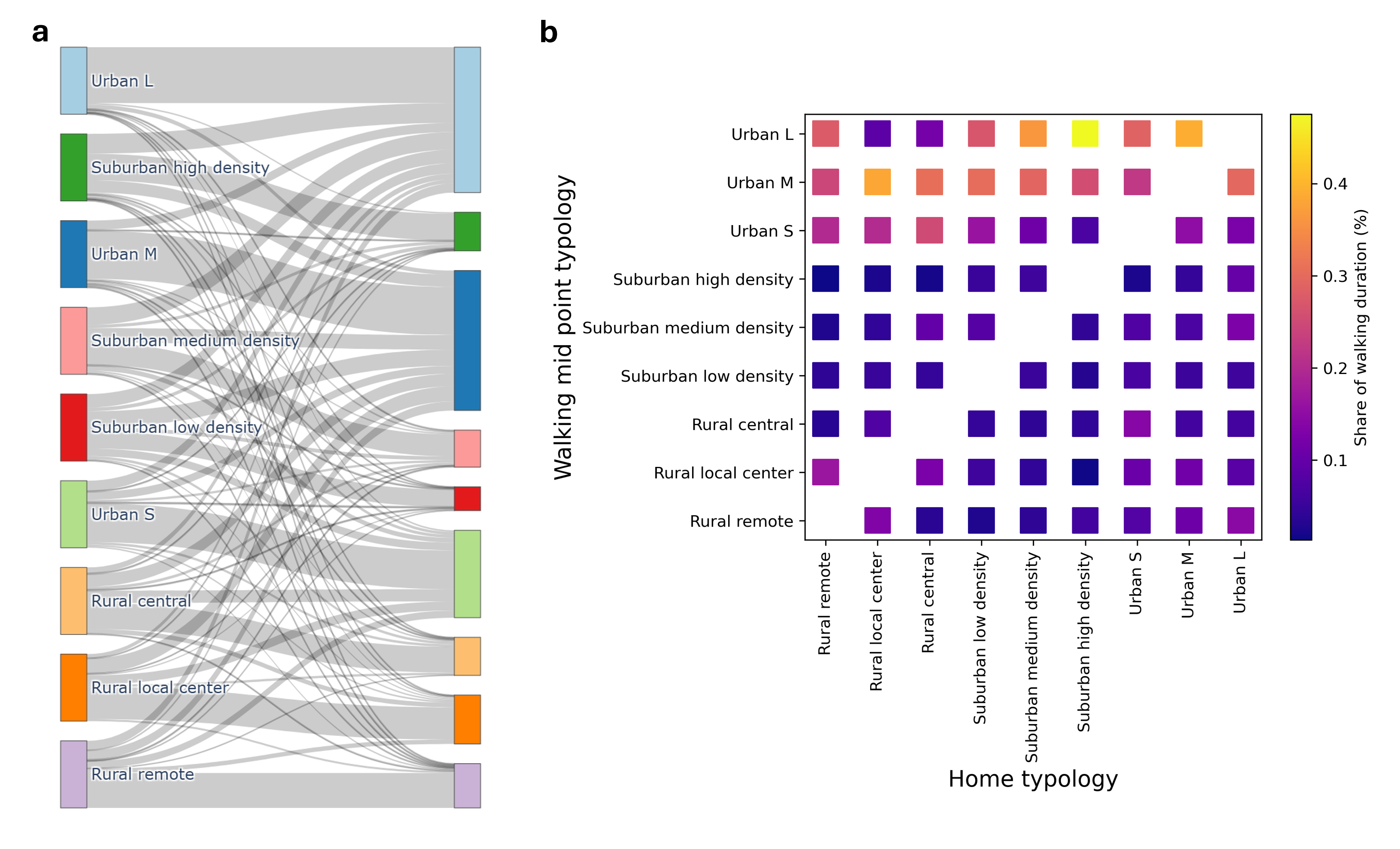}
\caption{ {\bf Home neighborhood typology and walking destination} (a) typology of residential municipality (left) to walking municipality (right)  (b) Share of walking duration for participants in each home typology at different walking destinations
}\label{sfig3}
\end{figure}

\begin{figure}[!h]
\centering
\includegraphics[width=0.9\textwidth]{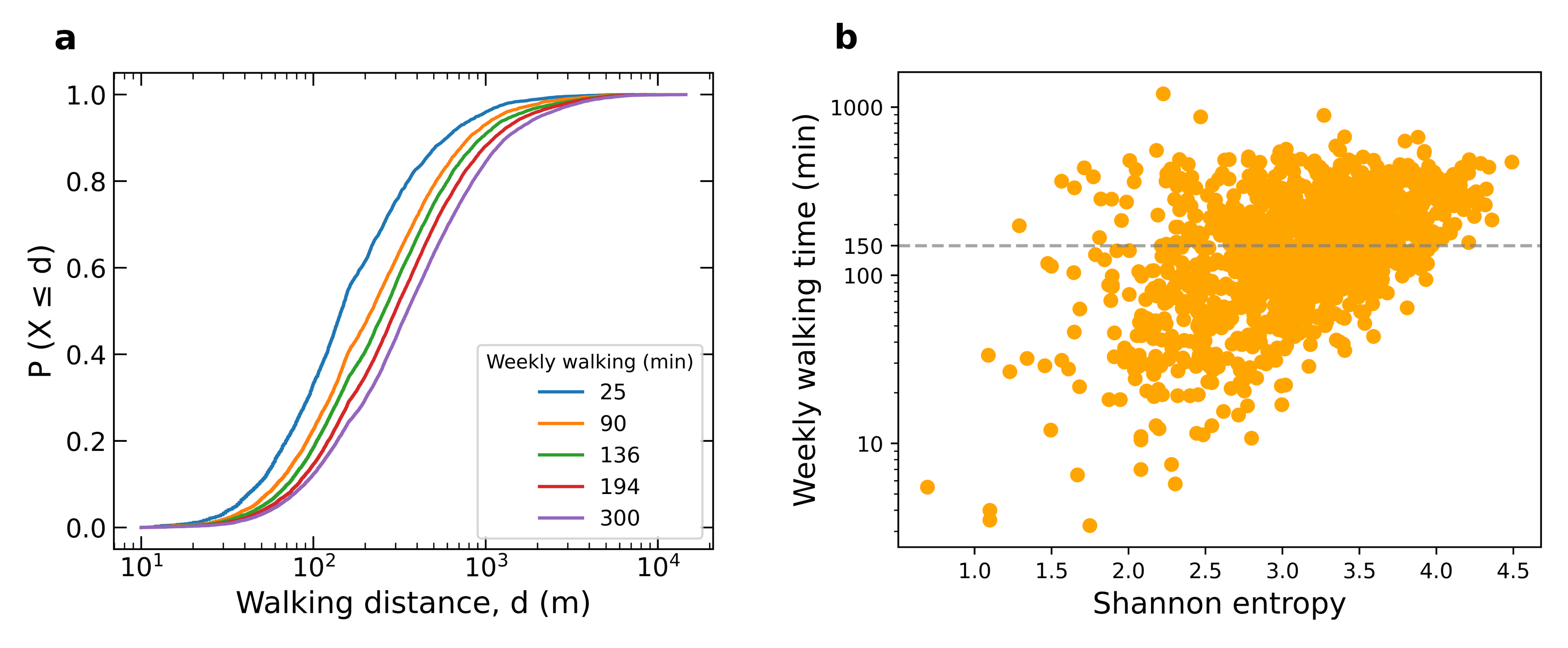}
\caption{ {\bf Variations in walking behaviors in TimeUse+.} (a) CDF of walking trip distance for different walking groups (b) Association between the Shannon entropy and walking time for each TimeUse+ participant
}\label{sfig4}
\end{figure}

\begin{table}[ht]
\centering
\caption{Summary statistics of travel distance by mobility mode}
\label{tab:event_distance}
\begin{tabular}{lrrr}
\hline
Statistic & Car & Cycle& Walk\\
\hline
Count& 63,058& 13,914& 117,166\\
Mean & 13,524.6& 3,568.3& 563.4\\
Std. & 20,448.1& 6,096.6& 855.2 \\
Min. & 25.0 & 11.0 & 10.0 \\
25\%& 2,539.0& 914.0 & 135.0 \\
50\% & 6,493.5& 1,664.5& 295.0 \\
75\% & 15,769.8& 3,703.0& 622.0 \\
Max. & 282,661.0& 122,427.0& 14,534.0\\
\hline
\end{tabular}
\end{table}

\end{document}